\documentclass[11pt]{article}
\usepackage{epsfig,amscd,amssymb}
\usepackage{graphicx}
\usepackage{latexsym}
\usepackage{amssymb,amsfonts}
\usepackage{amsmath} 
\begin{document}

%
%
%
%
\title{Hard squares with negative activity}%

\author{Paul Fendley$^1$, Kareljan Schoutens$^2$  and Hendrik van Eerten$^2$
\medskip \\ 
$^1$ Department of Physics, University of Virginia, \\
Charlottesville, VA 22904-4714 USA \\  
{\tt fendley@virginia.edu}\smallskip \\  
$^2$ Institute for Theoretical Physics,  
University of Amsterdam,\\
Valckenierstraat 65, 1018 XE Amsterdam, The Netherlands \\  
{\tt kjs@science.uva.nl, hveerten@science.uva.nl}}
\smallskip 

\date{August 23, 2004}

\maketitle

\begin{abstract}
We show that the hard-square lattice gas with activity $z=-1$ has a number
of remarkable properties. We conjecture that all the eigenvalues of
the transfer matrix are roots of unity. They fall into groups
(``strings'') evenly spaced around the unit
circle, which have interesting number-theoretic
properties. For example,
the partition function on an $M\times N$ lattice
with periodic boundary condition is identically 1 when $M$ and $N$ are
coprime.  We provide evidence for these conjectures from analytical and
numerical arguments.

\end{abstract}

\section{Introduction}

The hard-square model is a well-known model of two-dimensional
statistical mechanics \cite{square,BET}. It describes a
classical gas of particles on the square lattice, with the
restriction that particles may not be on adjacent sites. 
The activity $z$ is the Boltzmann weight per particle. One can
think of the particles as hard squares with area twice that of a
lattice plaquette. The squares are placed with their centers on a
lattice site and their corners at the four adjacent sites. The
restriction amounts to not allowing the squares to overlap, although
they can touch. The particles/squares do not interact, except 
via this hard core.  The partition function $Z$ for the hard
square model is then simply
\begin{equation}
Z = \sum_n A(n) z^n,
\end{equation}
where $A(n)$ is the number of allowed configurations with $n$
particles. Combinatorialists would describe $A(n)$ for an $M\times N$
lattice as the number of $M\times N$ matrices with $n$ entries $1$ and
the remaining zero, such that no row or column contains two
consecutive nonzero entries.

Despite its simple definition, the hard-square model has a number of
interesting properties \cite{Baxhard}. In this paper, we discuss the
fascinating structure present for the special case $z=-1$.  We will present
substantial analytic and numerical evidence that when $z=-1$, the
eigenvalues of the transfer matrix with periodic boundary conditions
are all roots of unity.  Moreover, for an $M\times N$ lattice with
periodic boundary conditions in both directions, we find $Z=1$ when
$M$ and $N$ are coprime.

There are a number of reasons why it is interesting to study the hard
square model at negative activity, even though the model has negative
Boltzmann weights. 

First, gases with hard cores are generically expected to exhibit a
phase transition in the Yang-Lee universality class at some negative
value of $z$ \cite{Fisher,CarYL}. For the hard-square model, this
transition takes place at $z=z_c\approx -.1$. Thus our $z=-1$ results
are describing the regime ``past'' this transition. This regime is
very poorly understood, even though it should be described by an
integrable field theory \cite{CM} (the Yang-Lee conformal field theory
with a perturbation of opposite sign than usual). In this regime there
are level crossings as $z$ is decreased past $z_c$ \cite{Yurov},
making a field-theory analysis difficult.

Second, the hard-square model is not known to be integrable for any
values of $z$ except for the trivial cases $z=0$ and $z=\infty$.
We thus do not know the origin of the behavior discovered here,
although the degeneracies of the levels and other results we describe
below do hint that there are symmetries yet to be uncovered. Hopefully
such symmetries will be useful in understanding the hard square model
for values of $z$ other than $-1$.

Third, gases with negative activity have been shown rigorously to be
equivalent to branched polymers in two dimensions higher \cite{BI},
and lattice animals in one dimension higher \cite{CarYL}.  More
precisely, the partition function of the lattice gas at negative $z$
is the generating function for branched-polymer configurations.

Fourth is that the partition function of $z=-1$ lattice gases arises
very naturally in the study of an interesting class of quantum models
with supersymmetry \cite{FS}. The simplest such model consists of
interacting fermions hopping on a lattice, subject to the constraint
that they cannot occupy adjacent sites. The partition function of the
{\it classical} hard-core model on the same lattice at $z=-1$ is the
Witten index of this quantum theory, and as such gives a lower bound
on the number of ground states of the theory. Thus the hard-square
partition function at $z=-1$ is the Witten index for this
supersymmetric model on the square lattice. We will discuss this and
other two-dimensional supersymmetric lattice fermions in a companion
paper \cite{FSnew}.

Finally, we believe that the results described in this paper alone
justify the study of this model: we know of no other non-trivial model
of two-dimensional statistical mechanics whose transfer matrix obey the
intriguing properties described below.  For example, the
hard-hexagon model (the analogous model on the triangular
lattice) is integrable, but numerical results indicate that for no
value of $z$ do its transfer-matrix eigenvalues become roots of
unity. 

In section 2 we define the transfer matrix used to obtain
these results. Readers interested in the results can safely skip most
of this and proceed to section 3, where our analytic and numeric
results for the partition function of the hard-square model at $z=-1$
are described.

\section{Transfer Matrix}

It is convenient to study the hard-square model by using its transfer
matrix.  For simplicity, we will consider periodic boundary
conditions, although much of what we say in this paper applies to open
boundary conditions as well. 

The dimension of the transfer matrix $D_N$ is the number of allowed
configurations along a circle with $N$ sites. We index these
configurations by an integer $i=1\dots D_N$; each configuration $i$
with $p$ particles is specified by the $p$ integers $(i_1,i_2,\dots
i_p)$, which give the locations of the particles in this
configuration.  The hard core means that $i_r\ne i_s,i_s\pm 1 ({\rm
mod} N)$ for any $r$ or $s$.  The number of configurations around a circle
is found by diagonalizing the transfer matrix
$\begin{pmatrix}
1&\sqrt{y}\cr
\sqrt{y}& 0
\end{pmatrix}$
for moving from one site to the next.
The contribution to $D_N$ of configurations
with $p$ particles is the coefficient of $y^p$ in
$$D_N (y)= \left(\frac{1+\sqrt{1+4y}}{2}\right)^N + 
\left(\frac{1-\sqrt{1+4y}}{2}\right)^N.$$
This generating function obeys the recursion relation
$$D_N(y) = D_{N-1}(y) + yD_{N-2}(y)$$ The total number of
configurations allowed with any $p$ is $D_N\equiv D_N(1)$. For
example, $D_1=1$ (coming from the configuration with no particles) and
$D_2=3$ (one configuration with no particles, and two with one
particle; because of the hard core two particles cannot be on two
consecutive lattice sites). Thus this sequence of $D_N$ goes as
1,3,4,7,11,18,\dots; see table 1 below. These are called Lucas
numbers, and obey the same recursion relation as do Fibonacci numbers.
If desired one can enumerate the configurations for any $N$
iteratively by this sort of approach. For example, the number of
lattice configurations with two consecutive fixed sites empty
$D^{00}_N$ are the Fibonacci numbers $1,1,2,3,5,\dots$; the number
with one of those two sites occupied is $D^{0x}_N=D^{00}_{N-1}$.

Consider two configurations $i$, $j$ having $p_i$ and $p_j$ particles
respectively.  These two configurations (each around a circle) are
allowed to be next to each other in the full two-dimensional model if
$i_r \ne j_s$ for all $r$ and $s$.  The transfer matrix $T_N$ acts on
a vector space $\mathbb{C}^{D_N}$; each basis element $v_i$ of this
vector space corresponds to a configuration $i$, and has $1$ in the
$i$th place and zeroes otherwise.  The partition function with
periodic boundary conditions in both directions is then
\begin{equation}
Z(M,N) = \hbox{tr}\,\left(T_N\right)^M.
\end{equation}
For the hard square model, 
if $i_r=j_s$ for some $r$ and $s$, then the transfer matrix entry
$T_{ij}=0$. If the two configurations are allowed next to each other,
then
\begin{equation}
(T_N)_{ij} = z^{(p_i+p_j)/2}\qquad\quad i\hbox{ allowed next to } j
\label{Tz}
\end{equation}
If $z$ is negative, by
convention we take the positive sign of the square root.
The transfer matrix is not unique, but will yield the same partition
function for any exponent $\lambda p_i + (1-\lambda)p_j$; the above
definition makes the matrix symmetric.  

Since the boundary conditions are periodic, the model has a
translation symmetry. The translation
generator acts on the vector space $\mathbb{C}^{D_N}$ as well, taking
a configuration with particles at $(i_i,i_2,\dots i_p)$ to the
configuration $(i_i+1,i_2+1,\dots i_p+1)$ where all
locations are interpreted mod $N$.  Thus ${\cal T}^N=1$,
and  the eigenvalues $t$ of ${\cal T}$ obey $t^N=1$.
It is easy to verify that $[T_N,{\cal T}]=0$. 
The transfer matrix therefore breaks into blocks
$T_N(t)$ acting on the eigenstates of ${\cal T}$ with eigenvalue $t$.  
Working in this basis makes
numerical computations more tractable, reducing the sizes of the
matrices involved by roughly a factor of $N$.  (Eventually this
doesn't help much, because $D_N$ grows exponentially.)

An eigenstate
of ${\cal T}$ with eigenvalue $t$ can be formed from each $v_i$ via
$${\cal V}_{[i]}(t) = (v_i + t^{-1}{\cal T}v_i + t^{-2}{\cal T}^2 v_i
+ \dots + t^{-(N-1)}{\cal T}^{N-1} v_i)\sqrt{{\cal N}_i}/N,$$ where
${\cal N}_i$ is the smallest integer which has ${\cal T}^{{\cal
N}_i}v_i =v_i$.  The normalization is chosen so that
${\cal V}^*_{[i]}\cdot {\cal V}_{[i]}=1$. For example, the state with
no particles has ${\cal N}_i=1$, while the state $i$ with particles at
$(2,4,\dots N)$ for $N$ even has ${\cal N}_i=2$.  The state ${\cal
V}_{[i]}(t)$ is non-zero only if $t^{{\cal N}_i}=1$.  Even though the
transfer matrix does not conserve the number of particles, ${\cal T}$
does, so the states ${\cal V}_{[i]}(t)$ have a fixed number of
particles $p_i$. Obviously any state $v_j$ which obeys $v_j = {\cal
T}^r v_i$ for some integer $r$ results in ${\cal V}_{[j]}= t^r {\cal
V}_{[i]}$.  Thus to give a complete (but not over-complete) set of
states we must choose just one particular $i$ in each of these sets.

Let us first examine the action of the transfer matrix in the sectors
with $t\ne 1$. The simplest state in these sectors is the one-particle
state, which we denote as $[1]$ (the translation-invariant state with
no particles is nonzero only in the $t=1$ sector). The transfer matrix
takes a one-particle state with a particle on the $r$th site to a
linear combination of all states which do {\it not} have a particle in
the $(r)$th place.  Using this, we see that the matrix element
$$ (T_N)_{[1][1]} = z(t+t^2+\dots t^N)=-z,$$
where we used $\sum_{k=0}^{N-1} t^k =0$.
For general matrix elements, we need the function
$\tau([i];t) = \sum_{r=1}^p t^{i_r}$
for each state $[i]$,
where as before the configuration $i$ has particles at $(i_1,i_2,\dots,
i_{p})$. Then
\begin{eqnarray*}
(T_N)_{[i][1]}(t\ne 1) = -z^{(p_i+1)/2} \sqrt{\frac{{\cal N}_i}{N}}\,
\tau([i];t)
\end{eqnarray*}
while $(T_N)_{[1][i]}(t) = (T_N)_{[i][1]}(1/t)$.  
Each term in these matrix elements arises
when configurations are forbidden to be next to each other. To get the
general matrix elements, the idea is likewise to see
which configurations are forbidden. The end result
is related to the product $\tau([i];t)\tau([j];1/t)$, but to not
overcount forbidden configurations, each term must have coefficient
$1$. Precisely, by using $t^N=1$ rewrite the product as
$$\tau([i];t)\tau([j];1/t) = \sum_{k=0}^{N-1} a_k([i],[j]) t^k.$$
Then we have
\begin{equation}
(T_N)_{[i][j]}(t\ne 1) = -z^{(p_i+p_j)/2} 
\frac{\sqrt{{\cal N}_i {\cal N}_j}}{N} \sum_{k=0}^{N-1} \theta(a_k([i],[j])) t^k
\end{equation}
where $\theta(a)=1-\delta_{a0}$, i.e.\ $\theta(0)=0$ and is $1$ otherwise.
Note that if $z$ is real and positive, $T_N$ is Hermitean.  In the
case of interest here $z$ is not, but if desired one can redefine $T_N(t)$
without changing its eigenvalues to make it symmetric.

Similar arguments give $T_N(t=1)$:
\begin{equation}
(T_N)_{[i][j]}(t= 1) = z^{(p_i+p_j)/2} 
\frac{\sqrt{{\cal N}_i {\cal N}_j}}{N} 
\left(N-\sum_{k=0}^{N-1} \theta(a_k([i],[j]))\right)
\end{equation}
where the $a_k$ are defined from $\tau([i];t)\tau([j];1/t)$ as above.
Note that when $[i]$ is the configuration with no particles, $\tau([i];t)=0$
and so all the corresponding $a_k=0$ as well.


\section{The partition function}

We write the partition function in terms of the roots
of the characteristic polynomial of the transfer matrix $T_N$
defined in the previous section.
The characteristic polynomial $P_N(x)$ is defined as
\begin{equation}
P_N(x) = \det(x - T_N).
\label{charpoly}
\end{equation}
The partition function for an $M\times N$ lattice with periodic
boundary conditions in both directions is
\begin{equation}
Z(M,N)= \sum_{i=1}^{D_N} (x_i(N))^M.
\label{part}
\end{equation}
where the $x_i$ are the roots of $P_N(x)$. 
Because $z$ is negative, the transfer matrix is not Hermitean;
a resulting complication we will discuss is that not all roots
$x_i$ need be eigenvalues of $T_N$.

Our main conjecture is that the roots $x_i(N)$ of the characteristic
polynomial are all roots of unity and for a given $N$ can be grouped
into ``strings''. A string is a set of $x_i$ evenly spaced around the
unit circle. We find just two kinds of strings, which we denote $S^+$
and $S^-$. The former are roots the values $x_k = e^{i2\pi k/S}$ for
$k=0,1,\dots S-1$, and the latter are the values $x_k = e^{i\pi
(2k+1)/S}$ for $k=0,1,\dots S-1$.  The existence of a string $S^{\pm}$
means that the polynomial $(x^S \mp 1)$ divides $P_N(x)$.  Moreover,
we find that all the values of $S$ for all strings for a given $N$
share a divisor with $N$, except for a single $1^+$ string for
every $N$.

The strong evidence for these conjectures comes from the numerical
results presented in Table 1. 
\begin{table}
\begin{center}
\begin{tabular}{|c|r|c|c|}
\hline
$N$ & $D_N$ &$P_N(x)$ \\
\hline \hline
1 & 1 & $x-1$  \\ 
2 & 3 &$(x^2+1)(x-1)$   \\ 
3 & 4 & $(x^3-1)(x-1) $ \\ 
4 & 7 & $(x^4-1)(x^2-1)(x-1) $ \\ 
5 & 11 &$(x^5+1)^2(x-1)  $ \\ 
6 & 18 &$(x^6-1)^2 (x^3-1)(x^2+1)(x-1) $ \\ 
7 & 29 & $(x^{14}+1)^2(x-1)   $ \\ 
8 & 47 & $(x^{10}-1)^4 (x^4-1)(x^2-1)(x-1) $ \\ 
9 & 76 & $(x^{18}-1)^2(x^9-1)^4  (x^3-1)(x-1)$ \\ 
10& 123 & $(x^{14}-1)^5(x^8-1)^5 (x^5-1)^2(x^2+1)(x-1)$ \\
11& 199 & $(x^{55}-1)^2(x^{22}+1)^4(x-1) $ \\
12& 322 & $(x^{24}-1)^2(x^{18}-1)^6(x^{12}-1)^{12}(x^6-1)^2 (x^4-1)
(x^3-1)(x^2-1)(x-1)$\\
13& 521 & $(x^{91}-1)^4(x^{26}-1)^4 (x^{13}+1)^4 (x-1)$\\
14& 843 & $(x^{28}-1)^4 (x^{22}-1)^7 (x^{16}-1)^{28} (x^{14}-1)^2 (x^{10}-1)^7
(x^7+1)^4 (x^2+1)(x-1)$\\
15& 1364 & $(x^{60}-1)^6 (x^{45}-1)^{18}(x^{15}-1)^{12}(x^5+1)^2
(x^3-1)(x-1)$\\
\hline
\end {tabular}
\end{center}
\caption{Characteristic polynomials of the transfer matrices $T_N$. The
roots of this polynomial give the partition function for an $M\times
N$ lattice with periodic boundary conditions, as in (\ref{part}).  }
\label{tab:dimcov}
\end{table}
We have checked up to $N=15$ that these conjectures hold.
Another result apparent
from this table is that det$(T_N)=1$, so $P_N(0)=(-1)^{D_N}$.
Additional numerical results for $N\le 9$ suggest that 
the roots of the characteristic polynomial for the transfer matrix
with open boundary conditions are also roots of unity.
For simplicity we will focus here on periodic.

The reason why the roots of $P_N(x)$ are all of unit modulus is
mystery to us; we do not know any other lattice gases even at $z=-1$
which share this property. 
Despite substantial effort, we have not found a formula for $P_N(x)$
for arbitrary $N$. 
However, one can exploit these conjectures to better understand the string
structure. We define the strings so that
only one kind of string is present for a
given $S$ and $N$. This means in $P_N(x)$ any occurrence of
$(x^S-1)(x^S+1)$ is combined into $(x^{2S}-1)$. 
Then we let $n_S^\pm(N)$ be the number of $S^\pm$ strings for a given $N$;
our convention means that $n_S^+(N)n_S^-(N)=0$.
Since each polynomial $P_N$ is of order $x^{D_N}$, we have
$$\sum_S S n_S^\pm (N)= D_N.$$ 
Another useful fact is that because $\sum_{k=1}^S
e^{2\pi i kM/S}=0$ unless $M$ is a multiple of $S$, only strings with
$M$ a multiple of $S$ contribute to $Z(M,N)$. An $S^+$
string contributes $S$, while an $S^-$ string contributes
$(-1)^{M/S}S$.

More about the string structure can be learned by exploiting modular
invariance, the symmetry of the model $Z(M,N)=Z(N,M)$
under interchange of the two cycles of the torus.  
A consequence is that knowing the $P_N(x)$ for
all the $N$ less than and equal to a given $S$ determines the numbers
of $S$-strings $n_S^\pm(N)$ for {\it all} $N$.

Let us explain how to implement this recursive procedure. Since $T_1
=1$, the only string for $N=1$ is a single $1^+$ string, as apparent
from Table 1. Thus $Z(M,1)=1$, and by symmetry $Z(1,N)=1$. Because
only strings where $M$ is a multiple of $S$ contribute to $Z(M,N)$,
only $1$-strings contribute to $Z(1,N)$. Since $Z(1,N)=1$, we have
just a single $1^+$ string: $n_1^+(N)=1$ and $n_1^-(N)=0$ for all $N$.
This is apparent in Table 1. Moving on to $N=2$, we have $D_2=3$ and
since there is exactly one 1-string, there must be just one
2-string. By our conjecture that $P_N(0)=(-1)^{D_N}$, this must be a
$2^+$-string, giving the result in the table. Thus
$$Z(2j-1,2)=1,\qquad\quad Z(4j-2,2)=-1, \qquad Z(4j,2) = 3,\quad j
\hbox{ an integer }\ge 1.$$ By symmetry, the same results hold for
$Z(2,N)$. Only $2$-strings and $1$-strings contribute to
$Z(2,N)$, and we already know there is exactly one $1^+$ string for all $N$. 
The explicit expression for $Z(2,N)$ then tells us that
$$n_2^+(4j) = 1, \qquad\qquad n_2^-(4j-2)=1 ,\qquad\qquad j \hbox{ 
an integer }\ge 1,$$
with all other $n_2=0$.  In a
similar fashion, one finds $n_3^+(3j) =1$ and $n_4^+(4j)=1$. These are all
apparent in table $1$.

For $N\ge 5$, this procedure yields only part of $P_N$ uniquely. For
$N=5,$ we know there is one $1^+$ string, and no $2,3$ or $4$
strings. Since $D_5=11$ and $P_5(0)=-1$, this means that there are
either two $5^+$ strings or two $5^-$ strings. We do not have a
general conjecture which will distinguish between the two
possibilities, so we must compute
$P_5(x)$ explicitly. As seen in table 1, there end up being two $5^-$
strings. Now we can use the symmetry under interchange of $M$ and $N$
to see that $n_5^+(10j)=2$ and $n_5^-(10j+5)=2$.  Likewise, to find
that there are two $6^+$ strings for $N=6$, one needs to find $P_6(x)$
explicitly. This then results in $n_6^+(6j)=2$. 
The general result is found by noting that for all strings with $S\ge
N$, $S n_S(N)$ is always a multiple of $N$.  This means that if we
were using a transfer matrix in the $M$-direction instead, this
contribution to the partition function 
would arise from $n_{N}(M)=Sn_S(N)/N$ strings of length
$N$. Using the results from
table 1, this procedure can be applied to find
$n_7^+(28j)=n_7^-(14(2j+1))=4$, $n_8^+(10j)=5$, $n_9^+(18j)=6$,
$n_9^+(9(2j+1))=4$, $n_{10}^+(14j)=7$, and so on.

Unfortunately, since $D_N$ increases exponentially with $N$, the
strings with $S<N$ make only a relatively small contribution to the
partition function.  As is apparent from the table, as $N$ increases,
the number of different types of strings with $S\ge N$ increases. We
have not yet seen any pattern to these numbers, but we are hopeful
that one may exist.  The increasing degeneracies (multiple roots at
the same $x$) apparent in the table as $N$ increases are an obvious
hint that there is some yet-undiscovered symmetry structure.

The conjecture that all values of $S$ for a given $N$
(except for a single $1^+$ string) share a divisor with $S$ means that
$$Z(M,N)=1 \qquad \hbox{when }M\hbox{ and }N \hbox{ are coprime.}$$
This fact is apparent in Table 2.  
\begin{table}
\begin{center}
\begin{small}
\begin{tabular}{r|rrrrrrrrrrrrrrrrrrrr}
 & 1 & 2 & 3 & 4 & 5 & 6 & 7 & 8 & 9 & 10 & 11 & 12 & 13 & 14 & 15 & 16 & 17 & 18 & 19 & 20\\
\hline
1  & 1 &  1 & 1 & 1 &  1 &  1 & 1 & 1 & 1 & 1 & 1 & 1 & 1 & 1 & 1 & 1 & 1 & 1 & 1 & 1 \\
2  & 1 & -1 & 1 & 3 &  1 & -1 & 1 & 3 & 1 & -1 & 1 & 3 & 1 & -1 & 1 & 3 & 1 & -1 & 1 & 3 \\
3  & 1 &  1 & 4 & 1 &  1 &  4 & 1 & 1 & 4 & 1 & 1 & 4 & 1 & 1 & 4 & 1 & 1 & 4 & 1 & 1 \\
4  & 1 &  3 & 1 & 7 &  1 &  3 & 1 & 7 & 1 & 3 & 1 & 7 & 1 & 3 & 1 & 7 & 1 & 3 & 1 & 7 \\
5  & 1 &  1 & 1 & 1 & -9 &  1 & 1 & 1 & 1 & 11 & 1 & 1 & 1 & 1 & -9 & 1 & 1 & 1 & 1 & 11 \\
6  & 1 & -1 & 4 & 3 &  1 & 14 & 1 & 3 & 4 & -1 & 1 & 18 & 1 & -1 & 4 & 3 & 1 & 14 & 1 & 3 \\
7  & 1 &  1 & 1 & 1 &  1 &  1 & 1 & 1 & 1 & 1 & 1 & 1 & 1 & -27 & 1 & 1 & 1 & 1 & 1 & 1 \\
8  & 1 &  3 & 1 & 7 &  1 &  3 & 1 & 7 & 1 & 43 & 1 & 7 & 1 & 3 & 1 & 7 & 1 & 3 & 1 & 47 \\
9  & 1 &  1 & 4 & 1 &  1 &  4 & 1 & 1 & 40 & 1 & 1 & 4 & 1 & 1 & 4 & 1 & 1 & 76 & 1 & 1 \\
10 & 1 & -1 & 1 & 3 & 11 & -1 & 1 & 43 & 1 & 9 & 1 & 3 & 1 & 69 & 11 & 43 & 1 & -1 & 1 & 13\\
11 & 1 & 1 & 1 & 1 & 1 & 1 & 1 & 1 & 1 & 1 & 1 & 1 & 1 & 1 & 1 & 1 & 1 & 1 & 1 & 1 \\
12 & 1 & 3 & 4 & 7 & 1 & 18 & 1 & 7 & 4 & 3 & 1 & 166 & 1 & 3 & 4 & 7 & 1 & 126 & 1 & 7 \\
13 & 1 & 1 & 1 & 1 & 1 & 1 & 1 & 1 & 1 & 1 & 1 & 1 & -51 & 1 & 1 & 1 & 1 & 1 & 1 & 1 \\
14 & 1 & -1 & 1 & 3 & 1 & -1 & -27 & 3 & 1 & 69 & 1 & 3 & 1 & 55 & 1 & 451 & 1 & -1 & 1 & 73 \\
15 & 1 & 1 & 4 & 1 & -9 & 4 & 1 & 1 & 4 & 11 & 1 & 4 & 1 & 1 & 174 & 1 & 1 & 4 & 1 & 11 \\
\end{tabular} \\
\end{small}
\end{center}
\caption{The partition function $Z(M,N)=Z(N,M)$ for $M\le 15, N\le 20$. }
\label{tab:two}
\end{table}
For values of $M$ and $N$ which are
not coprime, the partition function grows with increasing $M$ and $N$
at a much smaller rate as generic statistical mechanical systems. 
Since all the $x_i(N)$ have magnitude $1$, 
the maximal value of $Z(M,N)$ is the smaller of $D_N$ or $D_M$. 
$D_N$ grows
exponentially in $N$, while $Z(M,N)$ for generic values
of $z$ grows exponentially in $NM$.
Exponential growth in $NM$ is of course the standard behavior for
statistical-mechanical systems: it is the statement that the free
energy is extensive (a notable exception is systems with
supersymmetry). In fact even for the analogous systems at $z=-1$,
the partition function grows exponentially with the volume.  
For example, for the same model on the triangular lattice 
(the hard-hexagon model),
our numerics indicate that the partition function grows as
$Z_{\hbox{tri}}\approx (1.14)^{NM}$.


In this section we have been careful to refer to the $x_i(N)$ as the
roots of the characteristic polynomial $P_N(x)$, not as the
eigenvalues of $T_N$. The reason is that for $N\ge 4$, some of these
roots do not correspond to eigenvalues. The eigenvectors of a
non-Hermitean transfer matrix (even one of determinant 1 like $T_N$)
need not span the space of states
if two of the roots coincide. A simple example is the
matrix 
$A(b)=\begin{pmatrix}
1&b\\
0&1
\end{pmatrix}
$. Both roots are $1$, but for any $b\ne 0$ it has only
one eigenvector $e_1\equiv\begin{pmatrix} 1\\0\end{pmatrix}$.
The vector  $e_0\equiv
\begin{pmatrix} 0\\1\end{pmatrix}$ is linearly independent of $v_0$,
but obeys $A(b)e_0 = e_0 + (1+b)e_1.$ Acting repeatedly with $A$ does not change
the coefficient of $e_0$, but just continues to change the
coefficient of the eigenvector $e_1$. This is because matrices $A(b)$
and $A(b')$ commute; the associated conserved quantity is the
coefficient of $e_0$.

Despite this (interesting) complication, the partition function for
periodic boundary conditions in both directions can still be written
in the familiar form (\ref{part}); the sum is over all roots, not just
the eigenvalues.  For example, for $N=5$ and $t=1$, the transfer matrix is
$$T_5(t=1)=\begin{pmatrix}
1&i\sqrt{5}&\sqrt{5}\\
i\sqrt{5}&-4&3i\\
\sqrt{5}&3i&2
\end{pmatrix}$$
where the states are in order of increasing particle number $p_i=0,1,2$.
The characteristic polynomial of this is $(x-1)(x+1)^2$. There is only
one eigenstate of eigenvalue $-1$; the vector ${\cal
V}_0\equiv(0,1,i)$ is orthogonal to the eigenvectors but is
not an eigenvector itself.  Because the coefficient of ${\cal V}_0$ is
conserved as one acts with $T_5$, the partition function with periodic
boundary conditions remains (\ref{part}). This means that in this
sector the partition function is $Z(M,5)(t=1)=-1$ for $M$ odd, and
$Z(M,5)(t=1)=3$ for $M$ even.  However, the matrix elements of
$[T_5(t=1)]^2$ are not periodic in $M$ like this. The partition function
on a cylinder with fixed boundary conditions on the ends will not be
periodic; its magnitude will continue to increase, roughly linearly in $M$.

Moreover, one can use this to find a family of matrices with the same
roots $x_i$ as $T_5 (t=1)$. If all the roots $-1,-1,1$ were
eigenvalues, the matrix $[T_5(t=1)]^2$ would equal the identity. Here
$Q\equiv [T_5(t=1)]^2-1$ is non-zero; it annihilates the eigenvectors
and obeys $Q^2=0$.  The characteristic polynomial of $T_5(t=1)-\lambda
Q$ is independent of $\lambda$. When $\lambda=1/2$, this matrix has
three eigenvectors instead of two.  Analogous results follow for all
$N\ge 4$.  It is not clear, however, if the presence of this
nilpotent symmetry operator $Q$ will help in the analysis of the
model.

\bigskip\bigskip

This work was supported in part by the foundations FOM and
NWO of the Netherlands, by NSF grant DMR-0104799, and by
DOE under grant DEFG02-97ER41027.

\end{document}